\documentclass{article}
\usepackage{spconf,amsmath,graphicx}
\usepackage{amsthm}
\theoremstyle{plain}

\theoremstyle{definition}

\theoremstyle{remark}

\usepackage{algorithm}
\usepackage{algorithmicx}
\usepackage{algpseudocode}  
\usepackage{bm}
\usepackage{amssymb}
\usepackage{mathrsfs}
\newcommand{\vct}[1]{{\bm #1}}
\newcommand{\mtx}[1]{{\bm #1}}

\title{Scalable network adaptation for Cloud-RANs: An imitation learning approach\\
\thanks{This work was supported in part by the National Nature Science Foundation of China under Grant 61601290 and in part by the Shanghai Sailing Program under Grant 16YF1407700.} }
\name{Yifei Shen$^\dagger$, Yuanming Shi$^\star$, Jun Zhang$^\dagger$, Khaled B. Letaief$^{\dagger}$}
\address{$^\dagger$Dept. of ECE, The Hong Kong University of Science and Technology, Hong Kong\\
	$^\star$School of Information Science and Technology, ShanghaiTech University, Shanghai, China}
\begin{document}

\ninept
\maketitle
\begin{abstract}
Network adaptation is essential for the efficient operation of Cloud-RANs. Unfortunately, it leads to highly intractable mixed-integer nonlinear programming problems. Existing solutions typically rely on convex relaxation, which yield performance gaps that are difficult to quantify. Meanwhile, global optimization algorithms such as branch-and-bound can find optimal solutions but with prohibitive computational complexity. In this paper, to obtain near-optimal solutions at affordable complexity, we propose to approximate the branch-and-bound algorithm via machine learning. Specifically, the pruning procedure in branch-and-bound is formulated as a sequential decision problem, followed by learning the oracle's action via imitation learning. A unique advantage of this framework is that the training process only requires a small dataset, and it is scalable to problem instances with larger dimensions than the training setting. This is achieved by identifying and leveraging the problem-size independent features. Numerical simulations demonstrate that the learning based framework significantly outperforms competing methods, with computational complexity much lower than the traditional branch-and-bound algorithm. 
\end{abstract}
\begin{keywords}
Cloud-RAN, green communications, branch-and-bound, pruning, imitation learning.
\end{keywords}

\section{Introduction}
\label{sec:intro}

As mobile data traffic keeps growing exponentially, wireless networks are facing unprecedented pressure. Network densification is a promising way to further improve spectral and energy efficiency of wireless networks \cite{shi2018generalized}. However, it also imposes new challenges on interference management, radio source allocation and mobility management, as well as yielding high operating expenditure  \cite{peng2016recent}. Cloud radio access networks (Cloud-RANs) emerge as a cost-effective approach for densifying the network. It enables centralized signal processing by connecting the low-cost remote radio heads (RRHs) to the cloud data center via the optical fronthaul links. 

Network adaptation, e.g., adaptively switching off some RRHs to save power, is essential for the efficient operation of Cloud-RANs. It requires to optimize over discrete variables (i.e., the selection of RRHs and associated fronthaul links) and continuous variables (i.e. the downlink transmit beamforming vectors) \cite{shi2018generalized,shi2014group}. Unfortunately, it leads to mixed-integer nonlinear programming (MINLP) problems, which are highly intractable in general. Although global optimization algorithms such as branch-and-bound can find the globally optimal solution, the computational complexity is exponential in the worst-case. To alleviate the computational burden, heuristic algorithms based on convex relaxation have been recently proposed via exploiting the sparsity in the optimal solutions \cite{shi2014group} or simply relaxing the binary variables into the unit intervals \cite{cheng2013joint}. Despite good performance of these heuristics, the performance gaps are difficult to quantify or control. To obtain near-optimal solutions at affordable complexity, in this paper, we instead propose to approximate the branch-and-bound method to balance the computational complexity and solution gaps via machine learning.  

Inspired by the recent success of the ``learning to optimize'' paradigm \cite{li2016learning,he2014learning}, we propose to learn to prune in branch-and-bound for network
power minimization. This is motivated by the observation that the
computational complexity of branch-and-bound is mainly controlled by the pruning
policy. By formulating the pruning procedure as a sequential decision
problem, an imitation learning based training method is proposed. Data aggregation (DAgger) \cite{ross2011reduction} is further leveraged
to increase the precision of the solutions. To scale up to problem instances whose 
sizes are larger than that of the training instances, we propose to utilize problem-size independent
features for training. This is a unique advantage of the proposed framework and makes it scalable to larger network sizes. On top of that, our proposed framework only requires hundreds of samples by learning the policy at each node.
Numerical experiments demonstrate that this method
dramatically accelerates the branch-and-bound procedure, and significantly outperforms competing
methods. Its ability to scale to larger problem sizes is also demonstrated. Equipped with efficient convex optimization tools \cite{shi2015large}, the framework can be employed in real systems.


\section{System Model and Problem Formulation}\label{sec:sys_model}

\subsection{System Model}
Consider a Cloud-RAN with $L$ RRHs and $K$ single-antenna mobile users (MUs), where the $l$-th RRH is equipped with $N_l$ antennas. All the RRHs are connected to a baseband unit (BBU) pool via a high-bandwidth, low-latency fronthaul network, which performs centralized signal processing. We focus on coordinated downlink transmission, and consider the network power minimization problem. Let $\mathcal{L} = \{1,\cdots,L \}$ denote the set of RRH indices and $\mathcal{S}=\{1,\cdots,K  \}$ denote the index set of MUs. 

Assuming that all MUs employ single user detection, the corresponding signal-to-interference-plus-noise ratio (SINR) for the $k$-th MU is given by ${\rm SINR}_k = \frac{|\sum_{l \in \mathcal{L}}\vct{h}_{kl}^{\sf{H}}\vct{w}_{lk}|^2}{\sum_{i\neq k}|\sum_{l \in \mathcal{L}}\vct{h}_{kl}^{\sf{H}}\vct{w}_{li}|^2 + \sigma_k^2}, $ $\forall k \in \mathcal{S}$, where $\vct{w}_{lk} \in \mathbb{C}^{N_l}$ denotes the transmit beamforming vector from RRH $l$ to MU $k$, $\vct{h}_{kl} \in \mathbb{C}^{N_l}$ represents the channel vector between the $l$-th RRH and the $k$-th MU, and $\sigma_k^2$ is the variance of additive noise\cite{shi2014group}.


Let a binary vector $\vct{a}=(a_1,\cdots,a_L)$ with $a_i\in\{0,1\}$ denote the mode
of each RRH, i.e., $a_i = 1$ (resp. $a_i = 0$) if the $i$-th RRH and the corresponding
transport link are switched on (resp. switched off). Each RRH has its own transmit power constraint $\sum_{k\in\mathcal{S}} \|\vct{w}_{lk}\|_{\ell_2}^2 \leq a_l \cdot P_l, l \in \mathcal{L}$, where $\|\cdot\|_{\ell_2}$ is the $\ell_2$-norm of a vector.

\subsection{Problem Formulation}
The network power consumption in Cloud-RAN consists of the relative fronthaul network power consumption and the total transmit power consumption \cite{shi2014group}. Specifically, the relative fronthaul network power consumption is given by 
\begin{equation}
	f_1(\vct{a}) = \sum_{l \in \mathcal{L}} a_l\cdot P_l^c,
\end{equation}
where $P_l^c$ is the relative fronthaul link power consumption \cite{shi2014group}, i.e., the power saved when both the RRH and the corresponding fronthaul link are switched off.
The total transmit power consumption is given by 
\begin{equation}
	f_2(\vct{w}) = \sum_{l\in\mathcal{L}}\sum_{k\in\mathcal{S}} \frac{1}{\eta_l} \|\vct{w}_{k}\|_{\ell_2}^2,
\end{equation}
where $\eta_l$ is the drain efficiency of the radio frequency power amplifier, the aggregative beamforming vector $\vct{w}=[\vct{w}_1^T,\dots, \vct{w}_K^T]^T\in\mathbb{C}^{NK}$ with $\vct{w}_k = [\vct{w}_{1k}^T, \dots, \vct{w}_{Lk}^T]^T \in \mathbb{C}^{N}$ and $N=\sum_{l=1}^L N_l$.
%

Given SINR thresholds $\vct{\gamma}=(\gamma_1,\cdots,\gamma_K)$ for all the MUs and as an arbitrary phase rotation of a beamforming vector $\vct{w}_k$ does not affect SINR constraints, the SINR constraints can be expressed as a second order cone 
\begin{equation}
	\mathcal{C}(\vct{w}):\sqrt{\sum_{i \neq k}|\vct{h}_k^{\sf{H}}\vct{w}_i|^2 + \sigma_k^2 } \leq \frac{1}{\gamma_k}\Re(\vct{h}_k^{\sf{H}}\vct{w}_k), k \in \mathcal{S},
\end{equation}
where $\vct{h}_k = [\vct{h}_{1k}^T,\dots, \vct{h}_{Lk}^T]^T \in \mathbb{C}^{N}$, and $\Re(\cdot)$ denotes the real part of a complex scalar \cite{wiesel2006linear}.
The per-RRH constraints can be rewritten as 
\begin{equation}
\mathcal{G}(\vct{a},\vct{w}): \sqrt{\sum_{k\in\mathcal{S}} \|\mtx{A}_{lk}\vct{w}_{k}\|_{\ell_2}^2} \leq a_l \cdot \sqrt{P_l}, l \in \mathcal{L},
\end{equation}
where $\mtx{A}_{lk} \in \mathbb{C}^{N \times N}$ is a block diagonal matrix with identity matrix $\mtx{I}_{N_l}$ as the $l$-th main diagonal block matrix and zero elsewhere.

Hence, the network power consumption minimization problem can be formulated as the following MINLP problem:
\begin{equation}
        \begin{aligned}
        \mathscr{P} :&\underset{\vct{w},\vct{a}}{\text{minimize}}
        & & f_1(\vct{a}) + f_2(\vct{w})\\
        & \text{subject to}
        & & \mathcal{C}(\vct{w}),\mathcal{G}(\vct{a},\vct{w})\\
        & 
        & & a_l \in \{0,1\}, l \in \mathcal{L}.
        \end{aligned}
\end{equation}
Note that with a fixed binary vector $\vct{a}$, $\mathscr{P}$ is a second order cone programming (SOCP) problem.  This motivates a branch-and-bound approach \cite{lee2011mixed} to find a globally optimal solution.

\vspace{-0.5em}
\section{Global Optimization via Branch-and-Bound}
\vspace{-0.5em}

In this section, we shall present the standard branch-and-bound procedure to find a globally optimal solution for problem $\mathscr{P}$, followed by some observations.

\subsection{Branch-and-Bound}\label{sec:BnB}

Branch-and-bound algorithms \cite{lee2011mixed,balcan2018learning} build a binary search tree $\mathcal{T}$ iteratively. Each node of the tree contains a MINLP in the form of:
\begin{equation}
\begin{aligned}
\mathscr{P}_n(\mathcal{Z},\vct{z}) :&\underset{\vct{w},\vct{a}}{\text{minimize}}
& & f_1(\vct{a}) + f_2(\vct{w})\\
& \text{subject to}
& & \mathcal{C}(\vct{w}),\mathcal{G}(\vct{a},\vct{w})\\
&
& & \vct{a}_{[\mathcal{Z}]} = \vct{z} \\
& 
& & a_l \in \{0,1\}, l \in \mathcal{L},
\end{aligned}
\end{equation}
where $\mathcal{Z}$ is an index set, $\vct{a}_{[\mathcal{Z}]}$ is the elements of $\vct{a}$ indexed by $\mathcal{Z}$ and $\vct{z}$ is a given vector with $z_i\in\{0,1\}$. Its convex relaxation is given by:
\begin{equation}
\begin{aligned}
\mathscr{P}_R(\mathcal{Z},\vct{z}) :&\underset{\vct{w},\vct{a}}{\text{minimize}}
& & f_1(\vct{a}) + f_2(\vct{w})\\
& \text{subject to}
& & \mathcal{C}(\vct{w}),\mathcal{G}(\vct{a},\vct{w})\\
&
& & \vct{a}_{[\mathcal{Z}]} = \vct{z} \\
& 
& & 0 \leq a_l \leq 1, l \in \mathcal{L}.
\end{aligned}
\end{equation}

Branch-and-bound consists of three main components: a \emph{node selection policy}, a \emph{variable selection policy}, and a \emph{pruning policy}. At the beginning, $\mathcal{T}$ only consists of a root node containing MINLP $\mathscr{P}_n(\emptyset,\emptyset)$. At each iteration, the node selection policy selects a node containing MINLP $P$. Then a variable selection policy selects a variable $a_i$. Let $P^+_i$ (resp. $P^-_i$) denote problem $P$ with additional constraint $a_i=1$ (resp. $a_i=0$). Specifically, if $P$ is the same as $\mathscr{P}_n(\mathcal{Z},\vct{z})$, $P_i^+$ (resp. $P_i^-$) denotes the problem $\mathscr{P}_n(\mathcal{Z}\cup\{i\},[\vct{z},1])$ (resp. $\mathscr{P}_n(\mathcal{Z}\cup\{i\},[\vct{z},0])$). The right (resp. left) child of the node containing $P$ is assigned as a node containing $P_i^+$ (resp. $P_i^-$). Then branch-and-bound determines whether the node $P_i^+$(resp. $P_i^-$) is \emph{fathomed}. The node $P_i^+$ (resp. $P_i^-$) is fathomed if the optimal solution to the convex relaxation of $P_i^+$ (resp. $P_i^-$) satisfies the constraints in problem $\mathscr{P}$ or meets the pruning conditions. Iterations repeat until all the nodes are fathomed.

The node selection typically follows the depth first policy, the best first policy or the best estimation policy \cite{Conforti2014integer}. The variable selection policy mainly includes the most fractional policy \cite{achterberg2009scip}, the linear scoring policy \cite{linderoth1999computational}, the product scoring policy \cite{achterberg2009scip}, or the entropic lookahead policy \cite{2011information}.

The pruning policy is to remove nodes from the branch-and-bound tree to reduce the complexity and guarantee the global optimality of the returned solution. Let $P$ denote a given problem, $P_R$ denote its convex relaxation and $c^*_P$ denote the optimal objective value of $P_R$. We use $\mathcal{T}_P$ to represent the subtree whose root node is $P$. Then all the nodes in $\mathcal{T}_P$ can be removed from the binary search tree if one of the following situations holds: (1) $c^*_P > c^*$, where $c^*$ denotes the best solution satisfying constraints in $\mathscr{P}$ found ever; (2) $P_R$ is infeasible. Recall that $c^*_P$ provides a lower bound for all the problems in $\mathcal{T}_N$. If the lower bound is worse than the objective value of the best solution found, so does the original problem itself. Thus, the feasible set of all problems in $\mathcal{T}_N$ can not contain the optimal solution. Similarly, if $P_R$ is infeasible, all the problems in $\mathcal{T}_N$ must be infeasible. Therefore, removing the nodes in $\mathcal{T}_N$ will not affect the optimality.

\subsection{Observations}

Branch-and-bound is widely employed in solving MINLPs as it is capable to obtain the globally optimal solution. However, its computational complexity of is exponential, which can not be tolerated in many problems. As discussed above, the pruning policy is responsible for reducing the computational complexity. The more nodes are pruned, the less time we need to terminate the algorithm. Branch-and-bound guarantees the global optimality of the returned solution because it checks all other solutions are worse than the returned solution. In other words, most of the time is spent on checking non-optimal nodes. Therefore, pruning can be much more aggressive if we only want a promising solution rather than ensuring the optimality. This motivates us to learn a pruning policy via imitation learning, as will be described next. 
\vspace{-0.5em}
\section{Pruning via Imitation Learning}\label{sec:algo}
\vspace{-0.5em}



\vspace{-0.5em}
\subsection{Pruning as a Sequential Decision Problem}
In this section, we propose a framework to learn the pruning policy in branch-and-bound via imitation learning, which is instantiated on the framework to learn to search in mixed-integer linear programming \cite{he2014learning}.

Imitation learning consists of a sequential decision problem and an oracle \cite{daume2012course}. The sequential decision problem is defined by a state space $\mathcal{X}$, an action space $\mathcal{Q}$ and a policy space $\Pi$. A single trajectory consists of a sequence of states $x_1$,$\cdots$,$x_T$, a sequence of actions $q_1$,$\cdots$,$q_T$, and a policy $\pi \in \Pi$ that maps a state to an action $\pi(x_i) = q_i$. The oracle policy is a policy $\pi^*$ whose output action $q^*_i$ is always unquestionably sound. The key idea of imitation learning is to mimic an oracle's behavior based on the current state $x_i$. 

In branch-and-bound, the state $x_i$ consists of the problem data, the search tree visited, and the optimal solution and objective value of the relaxed problem at each visited node. The action is either to prune or not to prune a node. Thus, the action $q_i$ is a class in \{\emph{prune}, \emph{not prune}\} and the policy $\pi$ is a binary classifier. As for the oracle, the ideal one should achieve the optimal solution with the minimal number of nodes expanded. This condition holds if and only if we preserve the nodes whose feasible set contains the optimal solution of problem $\mathscr{P}$ and remove the others. For simplicity, we call these preserved nodes \textit{optimal nodes}. With oracle's policy, imitation learning can be reduced to a supervised learning problem. As we can hardly represent the state $x_i$, a feature mapping $\phi:\mathcal{X}\rightarrow \mathbb{R}^*$ is used to map the state $x_i$ into the feature vector $o_i$. If the dimension of features $o_i,i=1,\cdots,T$ is fixed and does not change with the problem dimension, we call them \emph{problem-size independent features}. The training examples are of the form $\{(o_1,q^*_1),\cdots,(o_T,q^*_T) \}$ such that observation $o_i \in \mathbb{R}^*$ is the feature vector and oracle's action $q^*_i$ is the label. The pruning policy, i.e., the classifier, attempts to learn a map from the feature vector to the oracle's action.

Ideally, we hope our learned policy to be able to handle any possible situations once it has been trained. Nevertheless, supervised learning might perform considerably badly when encountering a situation which is not recorded in the training dataset. DAgger emerges to address this issue, which is an iterative learning algorithm \cite{ross2011reduction}. Specifically, once we have learned a policy $\pi_i$, the pruning procedure is not controlled by the oracle but by $\pi_i$. Although the policy $\pi_i$ might make mistakes, we just let the mistakes happen and record the oracle's actions in these situations. A new dataset is generated according to the trajectory controlled by $\pi_i$ and the actions by the oracle, and a new policy $\pi_{i+1}$ is trained based on the new dataset. The new policy $\pi_{i+1}$ corrects the mistakes made by the old policy $\pi_i$.

Specifically, our algorithm consists of three stages: a training data generation stage, a training stage, and a testing stage. In the training data generation stage, we generate a training dataset $\mathcal{P}$ containing $|\mathcal{P}|$ problem instances. Branch-and-bound is used to find the globally optimal solution of each problem. Then we label the node on the path from the root to the optimal solution as \emph{not prune} and label the remaining nodes as \emph{prune}. In the training stage, an iterative training algorithm is used. At the $i$-th iteration, we have a trained policy (classifier) $\pi^{(i)}$. We extract the $i$-th problem from the training dataset. A standard branch-and-bound is performed to solve problem $\mathcal{P}_i$ except using classifier $\pi^{(i)}$ to prune nodes. $\pi^{(i)}$ might have some incorrectly classified instances. We collect these instances and their labels into set $\mathcal{D}$, and train a new classifier $\pi^{(i+1)}$ using data in $\mathcal{D}$. Such iteration repeats for $|\mathcal{P}|$ times and we return the policy $\pi^{(k)}$ that performs the best in the validation dataset. In the test stage, we just replace the standard pruning policy with the learned policy $\pi^{(k)}$ to solve problems. The pseudo-code of the iterative training algorithm is shown as Algorithm \ref{alg:A}.

\begin{algorithm}  
        \caption{Policy Learning ($\pi^*$)}  
        \label{alg:A}  
        \begin{algorithmic}
                \State $\pi^{(1)} = \pi^*$, $\mathcal{D}=\{\}$, $i \leftarrow 0$, $k \leftarrow 0$
                \For{$k = 1$ \textbf{to} $|\mathcal{P}|$ }
                \State $p \leftarrow \mathcal{P}_k$
                \State $\mathcal{N}=\{n_0\},\mathcal{D}^{(p)} = \{\}$
                \While{$\mathcal{N}\neq \emptyset$}
                \State $N \leftarrow$ select a node from $\mathcal{N}$
                \State $f \leftarrow \phi(N)$ 
                \If {$N$ is not fathomed}
                \If {$N \in \mathcal{N}_{opt}^{(p)}$ or $\pi^{(k)}(f)\neq prune$}
                \State $N_{i+1}^{(p)},N_{i+2}^{(p)} \leftarrow$ expand $N$
                \State $\mathcal{N} \leftarrow \mathcal{N} \cup \{N_{i+1}^{(p)},N_{i+2}^{(p)}\}$
                \State $i \leftarrow i+2$
                \EndIf
                \EndIf
                \If {$\pi^{(k)}(f) \neq \pi^*(f)$}
                \State $\mathcal{D}^{(p)} = \mathcal{D}^{(p)} \cup \{f,\pi^*(f)  \}$
                \EndIf
                \EndWhile
                \State $\mathcal{D} = \mathcal{D} \cup \mathcal{D}^{(p)}$
                \State $\pi^{(k+1)} \leftarrow$ train a classifier using data $\mathcal{D}$
                \EndFor
                
                \State \Return best $\pi^{(k)}$ on validation set
        \end{algorithmic}  
\end{algorithm}
In the algorithm, $\pi^*$ is the oracle's policy, $\mathcal{P}$ is the training problem dataset, and $n_0$ is the root node of the branch-and-bound search tree. $\mathcal{N}_{opt}^{(p)}$ is the set of optimal nodes. The expand operation first uses the variable selection policy to select a variable to branch on and then returns the two children of node $N$.

\vspace{-0.5em}
\subsection{Feature Design}

A good feature should be informative about both the problem itself and the branch-and-bound search tree. Moreover, it is also supposed to be friendly to classification algorithms and training data collection. Pruning a node with branching variable $a_i = 1$ (resp. $a_i=0$) implies that switching on (resp. switching off) the $i$-th RRH is not a sophisticated choice. Thus, the feature of a node whose branching variable is $a_i$ should at least convey the property of the $i$-th RRH and the corresponding fronthaul link. In addition, the learning algorithm utilizes the optimal solution of $\mathscr{P}$, whose computational cost grows exponentially as $L$ grows. Therefore, scaling up the algorithm to solve instances of much larger sizes than the training examples plays a pivotal role in accelerating the training process. As most classification algorithms can only deal with the situation where the dimension of input and output is fixed, employing problem-size independent features is critical for our algorithm to handle larger scale networks.

Specifically, suppose the current node containing problem $P$ and $a_i$ is the branching variable. The pruning policy is to determine whether to prune $P_i^+$ (resp. $P_i^-$). The feature includes four categories: (1) Problem features, i.e., partial data from the problem, which contain the relative fronthaul link power consumption and channel power gain of each RRH. Specifically, in order to let this kind of feature to be problem-size independent, the data we used should contain two parts: (a) the $i$-th relative fronthaul link power consumption divided by the summation of all the relative fronthaul link power consumption multiplied by the number of RRHs $\frac{L \cdot P^c_i}{\sum_{l \in \mathcal{L}} P^c_l}$. (b) the $i$-th RRH's channel power gain divided by the summation of all the channel power gains multiplied by the number of RRHs $\frac{L \cdot \sum_{k\in\mathcal{S}} \|\mtx{A}_{ik}\vct{h}_{k}\|_{\ell_2}^2}{\sum_{l \in \mathcal{L}} \sum_{k\in\mathcal{S}} \|\mtx{A}_{lk}\vct{h}_{k}\|_{\ell_2}^2}$. (2) Node features, computed merely from the current node $P_i^+$ (resp. $P_i^-$), which contain the depth, the plunge depth of $P_i^+$ (resp. $P_i^-$) and the optimal objective value $c^*_{P_i^+}$ (resp. $c^*_{P_i^-}$). (3) Branching features, computed from the branching variable $a_i$, which contain the value the branching variable $a^*_{P}[i]$. (4) Tree features, computed from the branch-and-bound search tree, which contain the optimal objective value at the root node, the number of solutions found ever, and the best objective value found ever $c^*$. We put problem features, optimal objective value, value of branching variable in a feature vector $\vct{o}\in \mathbb{R}^4$ and use it as the input of the classifier.

Due to the significant variations among the objective value of $\mathscr{P}$ under different network settings, all the objective values used as features in the branch-and-bound search tree should be normalized by the optimal objective value of the relaxed problem at the root node.

\subsection{Computational Analysis}\label{sec:format}
At each node, a relaxed SOCP problem needs to be solve, which is the main computation cost. Considering a network with $L$ RRHs and a node pruning policy which expands a non-optimal node with probability $\epsilon_1$ and prunes an optimal node with probability $\epsilon_2$, it can be shown that the expected number of SOCP to solve is $\mathcal{O}(L^2)$ when $\epsilon_1 \leq 0.5$ and $\mathcal{O}(L)$ when $\epsilon_1 \leq 0.3$. This demonstrates that the proposed framework enjoys a low expected computational complexity. 

\vspace{-0.5em}
\section{Numerical Experiments}
\vspace{-0.5em}
In this section, we present simulation results to compare our algorithm with some benchmark algorithms. The test dataset consists of $50$ network realizations with $L = 10$ 2-antenna RRHs and
$K = 15$ single-antenna MUs. The RRHs and MUs are uniformly and independently distributed in the square region $[−1000,1000] \times [−1000,1000]$. The fronthaul link power consumption is set to $P^c_l = (5+l)W, l=1,\cdots,10$. Other parameters are the same as in \cite{shi2014group}. 

In the first experiment, we generate $100$ network realizations for training and $50$ realizations for validation, which have the same number of RRHs, MUs, and $P^c_l, l=1,\cdots,10$, with the test dataset but different locations. For simplicity, the depth first policy is adopted as the node selection policy and the variable selection policy always selects the first unchosen variable. The classifier adopted here is support vector machine with the radial basis function kernel, which is implemented via libsvm \cite{CC01a}. The result is shown as ``Imitation Learning''. The first competing method is the relaxed mixed-integer nonlinear programming (RMINLP) \cite{cheng2013joint}, which turns off RRHs one by one based on the solution of relaxed MINLP. The second competing method is iterative group sparse beamforming (GSBF) \cite{shi2014group}, which leverages re-weighted $\ell_1$/$\ell_2$ norm to induce group sparsity to help select active RRHs.

We also test how the proposed framework scales up beyond the problem size in the training dataset. For the second experiment, in the training stage, we generate $200$ network realizations with parameters $L = 6$, $K = 8$, i.e. a smaller network size than the test set, and $P^c_l,l=1,\cdots,6$, uniformly distributed in $[6,15]W$. The validation dataset contains $50$ networks with parameters being the same as the test dataset. This result is shown as ``Scalable IL''.

                \vspace{-1em}
        \begin{figure}[htb]
                \centering
                \includegraphics[width=0.5\textwidth]{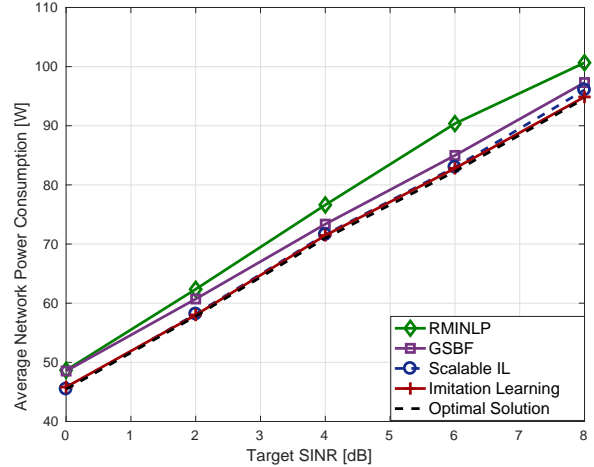}
                \caption{Average network power consumption versus TSINR.}
                \label{fig:learn_per}
        \end{figure}
        \vspace{-2.5em}

        \begin{table}[htb]
                
                \centering  
                \caption{Speedup and performance gap to branch-and-bound. The format is speedup/performance gap.} 
                \resizebox{0.5\textwidth}{!}{
                \begin{tabular}{|c|c|c|c|c|c|}  
                        \hline  
                        & TSINR=0 & TSINR=2 & TSINR=4 & TSINR=6 & TSINR=8\cr\hline
                        Imitation Learning&27.4x/0.06\%&21.0x/0.3\%&12.8x/0.6\%&3.6x/0.6\%&2.3x/0.1\%\cr\hline 
                        Scalable IL&14.2x/0.06\%&15.4x/0.6\%&7.8x/0.8\%&3.6x/0.8\%&1.8x/1.5\%\cr\hline 
                \end{tabular}}
                \label{tab:learn_per}
        \end{table}

The network power consumption is compared in Fig. \ref{fig:learn_per}, and the speedup and performance gap of the proposed framework are shown in Table \ref{tab:learn_per}. From Fig. \ref{fig:learn_per}, we see that the proposed framework not only significantly outperforms the competing methods, but also achieves near optimal results. With the same system size in both the training and test datasets, Table \ref{tab:learn_per} shows that the imitation learning based method speeds up standard branch-and-bound by a factor more than $20$ with the objective value loss less than $0.6\%$. ``Scalable IL'', which is trained on dataset with a system size much smaller than that of the test dataset, also achieves near-optimal performance with significant speedup over standard branch-and-bound. This shows that the proposed framework is capable to scale up to problem sizes beyond those of the training dataset. As the target SINR becomes larger, more RRHs must be turned on to ensure feasibility of the problem $\mathscr{P}$, and branch-and-bound becomes faster as the feasible search space is smaller. Therefore, the speedup of the proposed framework is less notable in this regime.

\vspace{-1em}
\section{Conclusions}
\vspace{-0.5em}

In this paper, we proposed an imitation learning based framework to learn to prune in branch-and-bound, which is applied to find a near-optimal solution for the network power minimization problem in Cloud-RANs. A unique advantage of this framework is that it is scalable to problem instances with different dimensions from those in the training dataset. This is achieved by identifying the problem-size independent features. The proposed framework is applicable to other MINLP problems in wireless networks such as the user admission control \cite{shi2016smoothed} and computation offloading problems \cite{mao2017survey}. 

\bibliographystyle{IEEEbib}
\bibliography{refs,strings}

\end{document}